\documentstyle[12pt,world_sci]{article}
\hbox{OITS-572}

\begin{document}

\title{{\bf HADRONIC PENGUINS IN $B$ DECAYS \\
AND EXTRACTION OF $\alpha$,
$\beta$ AND $\gamma$}}
\author{N.G. DESHPANDE
\thanks{Talk presented at the Beyond the Standard Model IV, Lake Tahoe,
California, December 13 - 18, 1994}
and XIAO-GANG HE\\
{\em Institute of Theoretical Science, University of Oregon\\
Eugene, OR 97403-5203, USA}}

\maketitle
\setlength{\baselineskip}{2.6ex}

\begin{center}
\parbox{13.0cm}
{\begin{center} ABSTRACT \end{center}
{\small \hspace*{0.3cm} Electroweak penguin operators become important when
$m_t \approx 174$ GeV. We discuss their implications for measuring CP violating
phases
$\alpha$, $\beta$ and $\gamma$.}}
\end{center}
\section{Introduction}
In this talk we shall focus on the isospin properties of penguin operators and
their
consequences for determination of CP violating phases $\alpha$, $\beta$
and $\gamma$. The relatively large electroweak contribution to the penguin
operators
resulting from the large t quark mass plays an important role\cite{1}.

The Hamiltonian for the charmless process $b \rightarrow s \bar q q$ in the
next to
leading order is given by\cite{2,3}
\begin{eqnarray}
H_{\Delta B=1} = {G_F\over \sqrt{2}}[V_{ub}V^*_{us}(c_1O^u_1 + c_2 O^u_2)
 - V_{tb}V^*_{ts}\sum^{10}_{i=3} c_iO_i] +H.C.\;,
\end{eqnarray}
where $c_i$'s are the Wilson coefficients defined at $\mu \approx m_b$ and the
operators $O_i$ are
defined as
\begin{eqnarray}
O^u_1 = \bar s_\alpha \gamma_\mu L u_\beta\bar
u_\beta\gamma^\mu L b_\alpha\;&,&\;\;\;\;
O^u_2 = \bar s \gamma_\mu L u\bar
u\gamma^\mu L b\;,\nonumber\\
O_{3,5} = \bar s \gamma_\mu L b \sum_{q'}
\bar q' \gamma_\mu L(R) q'\;&,&\;\;\;\;
Q_{4,6} = \bar s_\alpha \gamma_\mu L b_\beta \sum_{q'}
\bar q'_\beta \gamma_\mu L(R) q'_\alpha\;,\nonumber\\
O_{7,9} ={3\over 2}\bar s \gamma_\mu L b \sum_{q'} e_{q'}\bar q'
\gamma^\mu R(L) q'\;&,&\;\;
Q_{8,10} = {3\over 2}\bar s_\alpha \gamma_\mu L b_\beta \sum_{q'}
e_{q'}\bar q'_\beta \gamma_\mu R(L) q'_\alpha\;,\nonumber\\
\end{eqnarray}
with $L = 1-\gamma_5$ and $R = 1+\gamma_5$.  For $m_t = 174$ GeV, we
have\cite{3}
\begin{eqnarray}
c_1 &=& -0.3125\;, \;\;c_2 = 1.1502\;, \;\;c_3 = 0.0174\;, \;\;c_4 = -0.0373\;,
\nonumber\\
c_5 &=& 0.0104\;,
\;\;c_6 = -0.0459\;,\;\; c_7 = -1.050\times 10^{-5}\;,\nonumber\\
 c_8 &=& 3.839\times 10^{-4}\;,
\;\;c_9 = -0.0101\;, \;\;c_{10} = 1.959\times 10^{-3}\;.
\end{eqnarray}

Note that the tree operators $O_{1,2}$ transform as $\Delta I = 0,1$. The gluon
induced
penguin operators $O_{3-6}$ transform as $\Delta I = 0$ and the electroweak
penguin operators
$O_{7-10}$ again transform as $\Delta I = 0,1$. The largeness of coefficients
$c_9$ makes the
separation of tree and penguin extremly difficult based on iso-spin
transformation properties.
Similar remarks also apply to $ b\rightarrow d \bar q q$ process where tree
operators have
$\Delta I = 1/2$, $3/2$, gluon induced penguins are $\Delta I = 1/2$ and
electroweak penguins
have $\Delta I = 1/2$ and $3/2$.

\section{Probelm with measuring $\gamma$ through GLR proposal}

This is a proposal made by Gronau, London and Rosner\cite{4}. In this proposoal
the idea is to
exploit relations between $B^+\rightarrow K^0 \pi^+$, $B^+\rightarrow K^+\pi^0$
and
$B^+\rightarrow \pi^+\pi^0$ to extract $\gamma = \mbox{arg}(V_{ub}^*)$.

Using iso-spin decomposition, one can deduce
\begin{eqnarray}
A(B^+\rightarrow K^0\pi^+) + \sqrt{2} A(B^+\rightarrow K^+\pi^0) = \sqrt{3}
A_{3/2}\;.
\end{eqnarray}
If electroweak penguin is neglected, $A_{3/2}$ would be purely tree process and
its phase would
be given by $\gamma$. Further, using SU(3) symmetry, the magnitude of $A_{3/2}$
could be related
to $B^+\rightarrow \pi^+\pi^0$
\begin{eqnarray}
A_{3/2} = \sqrt{{2\over 3}} {V_{us}\over V_{ud}} A(B^+\rightarrow
\pi^+\pi^0)\;.
\end{eqnarray}
Considering similar relations for anti-particles, and assuming
\begin{eqnarray}
|A(B^+\rightarrow K^0\pi^+)| = |A(B^-\rightarrow \bar K^0 \pi^-)|\;.
\end{eqnarray}
a relationship true in factorization approximation, the phase $\gamma$ can be
determined by
constructing the two triangles for $B^+$ and $B^-$ decays given by Eq.(4).

If electroweak penguins are included, one finds in the  factorization
approximation\cite{1}
\begin{eqnarray}
A_{3/2} = {G_F\over \sqrt{2}} |V_{ub}^*V_{us}| (-1.05 e^{i\gamma + i\delta_T}
+0.84 e^{i\delta_P}) GeV^3\;,
\end{eqnarray}
where the first contribution is from the tree operators while the second
contribution is from
electroweak penguin operators, and $\delta_{T,P}$ are undetermined strong
rescattering phases.
The largeness of penguin contribuiton makes it impossible to relate $A_{3/2}$
to $B^+\rightarrow
\pi^+\pi^0$ process which is dominantly a tree process, and therefore the
method fails.

\section{Measurement of $\alpha$ and $\beta$}

CP asymmetry measurement $B^0\rightarrow f_{CP}$ versus $\bar B^0\rightarrow
f_{CP}$ where $f_{CP}$
is CP eigenstate like $\psi K_S$, $\pi^+\pi^-$, ... is a prefered way of
measuring angle $\alpha$
and $\beta$\cite{5}. We define
\begin{eqnarray}
A = <f_{CP}|H|B^0>\;,\;\; \bar A = <f_{CP}|H|\bar B^0>\;,
\end{eqnarray}
and mass-mixing parameter
\begin{eqnarray}
{q\over p} = \sqrt{{M_{12}^*\over M_{12}}}\;.
\end{eqnarray}
Introduce $\lambda = (q/p)(\bar A/ A)$, then if $|\lambda| = 1$,
\begin{eqnarray}
A_{asy} = - \mbox{Im}(\lambda) \mbox{sin}(\Delta m t)\;.
\end{eqnarray}
In the Standard Model $\mbox{arg} (q/p) = -2\beta$. For $f_{CP} = \psi K_s$,
$\bar A/A = 1$ and the
asymmetry is a measurement of $\mbox{sin} (2\beta)$.

For $f_{CP} = \pi^+\pi^-$, if tree amplitude dominates,
$\bar A /A = e^{-i2\beta}$
and $\mbox{Im}\lambda = - \mbox{sin}2(\beta +\gamma) = \mbox{sin}(2\alpha)$.
However, the penguin
contribuitons are quite significant with electroweak penguin contributing about
one third of the
total. We find for the ratio of penguin to tree operators in the
amplitude\cite{1}
\begin{eqnarray}
R_{\pi^+\pi^-} &=& \left |{A_P(B^0\rightarrow \pi^+\pi^-)\over
A_T(B^0\rightarrow \pi^+\pi^-)}\right |
 = 0.07 \left |{V_{td}\over V_{ub}}\right |\;,\nonumber\\
R_{\pi^0\pi^0} &=& \left |{A_P(B^0\rightarrow \pi^0\pi^0)\over
A_T(B^0\rightarrow \pi^0\pi^0)}\right |
= 0.23 \left |{V_{td}\over V_{ub}}\right |\;.
\end{eqnarray}
The error in $\mbox{Im}(\lambda)$ is given by
\begin{eqnarray}
\mbox{Im}(\lambda) - \mbox{sin}(2\alpha) = -R_{\pi\pi}
{2\mbox{cos}(\delta)\mbox{sin}(\alpha) + \mbox{sin}(2\alpha) (R_{\pi\pi}
-2 cos(\delta +\alpha)\over 1 + R^2_{\pi\pi} - 2 R_{\pi\pi}
\mbox{cos}(\delta+\alpha)}\;,
\end{eqnarray}
where $\delta$ is a unknown strong rescattering phase.
If $\alpha = 45^0$, the error in determination of $\alpha$ could be as large as
$13^0$ for $f_{CP} =
\pi^+\pi^-$. For $f_{CP} = \pi^0\pi^0$ the uncertainty is even larger.

We further find that rate of $B^0\rightarrow \pi^0\pi^0$ is quite small, so the
method proposed by
Gronau and London\cite{6} to disentagle penguin contamination seems hard to
implement.

In conclusion, we find that the electroweak penguins have large effect on the
determination of the CP violating phase $\gamma$ and also non-negligible effect
on measuring $\alpha$. Care must be taken when interpreting the experimental
data. The determination
of $\beta$ is not affected by the electroweak penguins.

\bibliographystyle{unsrt}

\end{document}